\documentclass[twocolumn]{aastex63}

\usepackage{hyperref}
\usepackage{verbatim}
\graphicspath{{./}{figures/}}
\received{2019 December 4}
\revised{2020 February 3}
\accepted{2020 February 27}
\usepackage{amsmath}
\usepackage{amsfonts}
\shorttitle{Winds on Black Widow Companions}
\shortauthors{Kandel \& Romani}
\defcitealias{linares2018peering}{LSC18} 	
\defcitealias{romani2015keck}{RGFK15} 	
\defcitealias{van2011evidence}{VBK11}

\begin{document}

\title{Atmospheric Circulation on Black Widow Companions}

\correspondingauthor{D. Kandel}
\email{dkandel@stanford.edu}

\author[0000-0002-5402-3107]{D. Kandel}
\affil{Department  of  Physics,  Stanford  University,  Stanford,  CA, 94305, USA}

\author[0000-0001-6711-3286]{Roger W. Romani}
\affil{Department  of  Physics,  Stanford  University,  Stanford,  CA, 94305, USA}
\begin{abstract}

We present a model for atmospheric wind circulation in binary millisecond pulsar (MSP) companions, showing how the optical light curve (LC) and radial velocities (RV) are sensitive to the wind flow, causing LC orbital phase shifts and asymmetries, as observed for several `spider' MSP. Velocity widths of spectral lines offer additional opportunities for measuring surface wind speed. As examples, we fit optical data for a black widow pulsar J1959+2048 and a redback pulsar J2215+5135; the wind heating models (WH) are statistically strongly preferred over direct heating (DH) for both objects, although the latter is even better fit with a heated spot. In general, WH effects tend to increase the inferred orbital inclination $i$ and decrease the inferred companion center-of-mass radial velocity amplitude $K_\mathrm{c}$; both effects decrease the inferred neutron star mass. Even with such a decrease, we find large masses for the two neutron stars: $2.18\pm0.09M_\odot$ and $2.28^{+0.10}_{-0.09}M_\odot$, respectively (for the modest surface speeds fit from the bulk heat flow; supersonic photospheric winds can slightly change these values). These are among the highest masses known, and our improved modeling increases confidence that the results are important for understanding the dense matter equation of state. 
\end{abstract}


\section{Introduction} \label{sec:intro}
Black widows (BWs) and redbacks (RBs), together known as  `spiders',  are binary systems of millisecond pulsars (MSPs) and low-mass companions in tight $<1$\,d orbits, with the companion heated and evaporated by the pulsar spindown power. Since the discovery of BW PSR J1959+2048's companion, it has been known that the heating pattern and the resulting light curve are important probes of the binary geometry (\citealt{djorgovski1988photometry}, \citealt{callanan1995orbital}) and the mass of the MSP (\citealt{aldcroft1992spectroscopy}, \citealt{van2011evidence}). With the advent of the {\it Fermi} LAT sky survey and attendant follow-up searches, the number of known `spiders' has greatly increased. We now have detailed light curves of many of these objects \citep{draghis2019multiband}. Standard light curve modeling assumes direct pulsar irradiation of the companion, which then re-emits thermal radiation. Thus, in this picture, the optical light curve is symmetric about a maximum at pulsar inferior conjunction.

While rough measurements of the optical modulation are broadly consistent with this picture, more detailed light curves often show significant peak asymmetries \citep{stappers2001intrinsic} and phase-shifts \citep{schroeder2014observations}. Several elaborations to the heating model have been introduced to explain such effects, including companion irradiation by an asymmetric intrabinary shock (IBS; \citealt{2016ApJ...828....7R}) or propagation of shock particles to heat a companion magnetic pole \citep{2017ApJ...845...42S}. Here we discuss another effect, companion super-winds, that can advect the pulsar heat and affect the light curves and phase-resolved spectroscopy. Such winds have a direct analog in similar surface circulation inferred for the so-called `hot Jupiters'. 

 In this paper, we present a model for global wind circulation in spider companions, starting in \S \ref{model} from simple analytic expressions that illustrate the principal effects, without following the detailed hydrodynamics. We describe an efficient numerical realization in the ICARUS light curve fitting code (\S 3), and include a treatment of the modification of the companion radial velocity curve. We next apply this model to two spider pulsars (\S \ref{application}). The results certainly show dramatic improvement over direct heating fits, but since other models (e.g. the heated magnetic pole described above) can induce asymmetry we compare with a simple hot-spot model. In one BW example, the wind model is mildly preferred; for the RB example, the spot model provides a better fit. Interestingly both models give very similar binary parameters, including inclination and mass, implying that these fits can be robust to the physical details of the heating. However, we conclude by describing the observational improvements needed to distinguish the physical situation.
 
\section{Wind Heat Redistribution Model}\label{model}
There is increasing evidence that many hot Jupiters display winds with a strong equatorial eastward flow  (\citealt{showman2002atmospheric}, \citealt{seager2005dayside}; see \citealt{hs15} for a recent review). We might expect similar winds for the spider companions. Although exoplanets can have similar orbital periods (e.g. 4.3h for K2-137b; \citealt{smith2017k2}) and stellar temperatures (e.g. $\sim3400$\,K for WASP-33b; \citealt{2015A&A...584A..75V}), spider companions have relatively high masses, strong sub-photospheric heating and short orbital periods (strong Coriolis effects), so we might expect the global wind effects to be even stronger for these objects. Also, since the {\it only} significant visible light source in these systems is the heated companion, optical photometry and spectra can probe these effects more directly than possible for hot Jupiters. These factors make the spiders interesting for the probe of global circulation effects.

\begin{figure}
    \centering
    \includegraphics[scale=0.5]{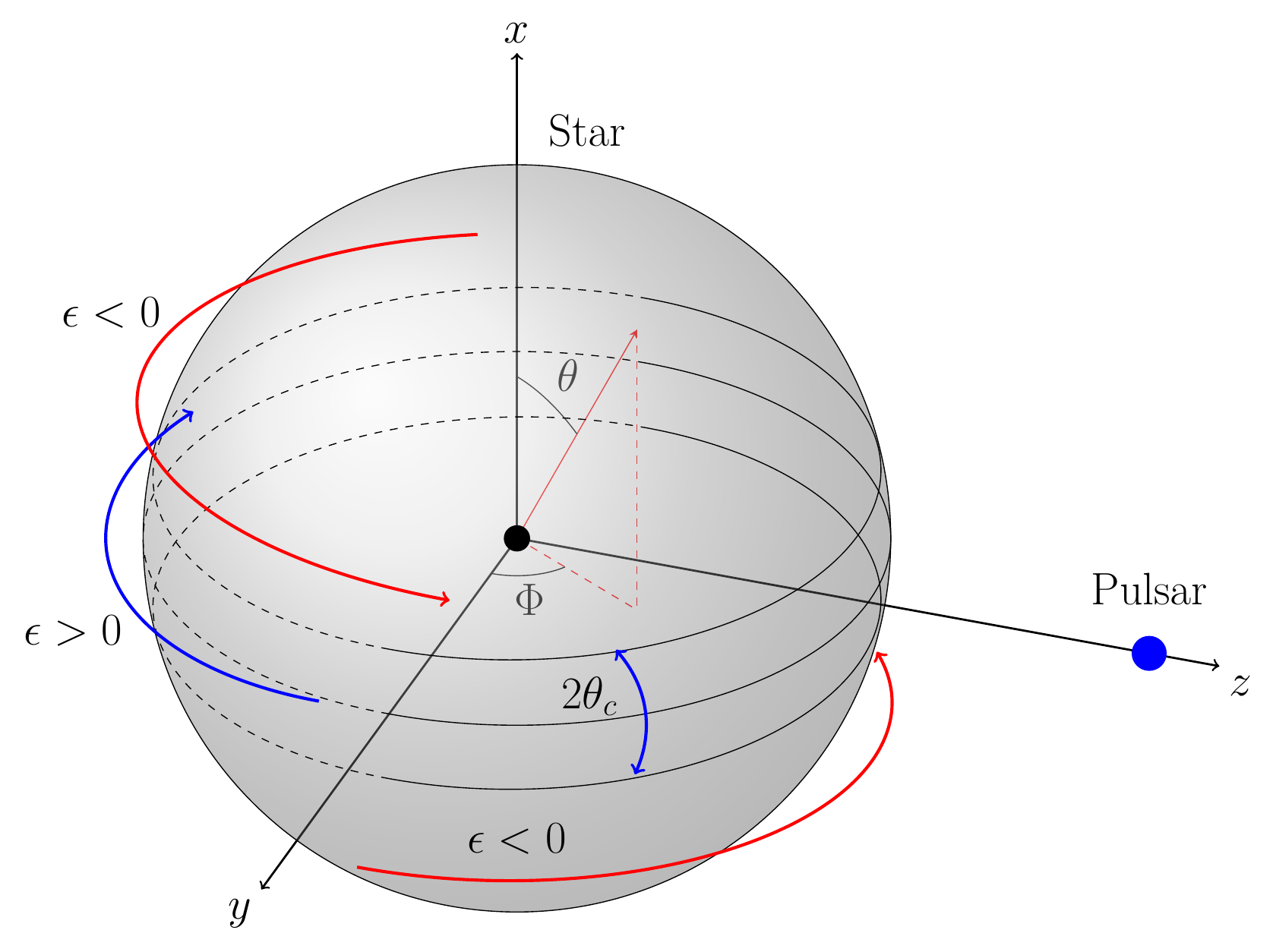}
    \caption{Schematic of the idealized wind flow and coordinate system used.}
    \label{fig:schematic}
\end{figure}

In a simple heuristic model, we assume that a tidally locked companion is heated on the dayside by pulsar irradiation (with a given parcel receiving a flux $\propto \cos\Psi$, with $\Psi$ the angle between the local normal and the direction to the pulsar). The interior also provides heat which produces a characteristic un-irradiated night side temperature $T_N$. Finally, the heated gas is assumed to flow along the line of constant companion latitude $\theta$ (in the E-W direction) which transports heat from the point of absorption. A gas parcel re-emits this heat at some temperature $T$. We normalize all temperatures to a characteristic $T_0\equiv (L_{\rm{P}}/4\pi\sigma a^2)^{1/4}$ at the sub-pulsar point, with $L_{\rm{P}}$ the isotropically radiated pulsar luminosity and $a$ the orbital separation. This results in a dimensionless energy balance equation for the companion temperature profile $\tilde{T}\equiv T/T_0$ along a given latitude (see \citealt{cowan2010model})
\begin{equation}\label{eq:kinematic}
    \frac{d\tilde{T}}{d\Phi} = \frac{1}{\epsilon}\left(\tilde{f}_{\mathrm{P}}(\Phi) + \tilde{T}_N^4 - \tilde{T}^4\right)~,
\end{equation}
where $\Phi$ is the azimuth angle from the sub-pulsar point, $f_p(\Phi)$ is the pulsar flux at this angle, and $\epsilon\equiv \tau_{\rm{rad}} \omega_{\rm{adv}}$ is the ratio of radiation time to advection time. For strict tidal locking, $\epsilon = 0$, meaning no wind energy transport across the star. Super-rotating (Eastward-going) winds have $\epsilon>0$ while sub-rotating winds have $\epsilon<0$. Numerical simulations of planetary atmospheres show that often the wind at equatorial region moves from the sub-stellar point in eastward direction (blue arrow in Fig. \ref{fig:schematic}), whereas the wind at mid-latitudes extending up to poles move in the westward direction (see e.g. \citealt{cowan2010model}). While the detailed physics is not fully understood, this zonal global wind structure can be traced to the asymmetry of the Coriolis forces \citep{dobbs2008atmospheric}.

\subsection{Analytic Approximate Model }\label{sec:toy_application}

Here we describe the asymptotic behaviour of Eq. \eqref{eq:kinematic}, and numerically compute the temperature profile for a simple spherical symmetrical companion and isotropic pulsar flux. In the limit of $\epsilon\rightarrow 0$, the term $\epsilon\, \rm{d}T/\rm{d}\Phi\rightarrow 0$, thus Eq. \eqref{eq:kinematic} simply reduces to the temperature profile corresponding to direct heating
\begin{equation}
    T^4\approx f_p(\Phi) + T_N^4~,
\end{equation}
In the opposite limit when $\epsilon\gg 1$, the left-hand term in Eq. \eqref{eq:kinematic} vanishes, thus giving
\begin{equation}
    T(\Phi) \approx \text{constant}~.
\end{equation}

\begin{figure}
    \centering
    \includegraphics[scale=0.42]{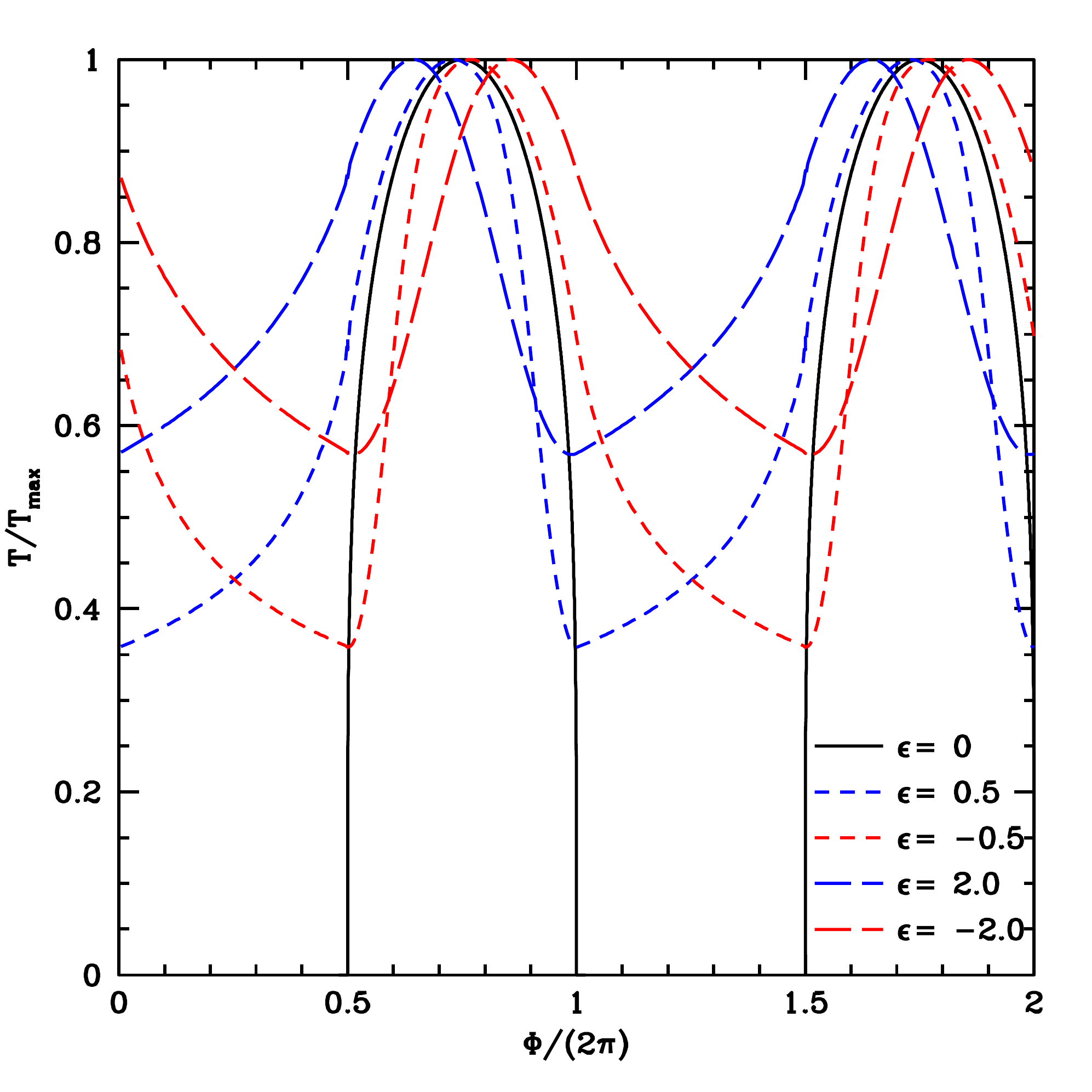}
    \caption{Temperature of the star as a function of the azimuth phase.}
    \label{fig:wind_model}
\end{figure}

For a spherical companion the flux from pulsar at fixed angle $\theta$ is $f_P(\Phi)\propto \max(\cos\Phi, 0)$, and Eq. \eqref{eq:kinematic} becomes
\begin{equation}
    \frac{dT}{d\Phi} = \frac{1}{\epsilon}\left(\max(\cos\Phi, 0) + T_N^4 - T^4\right)~.
\end{equation}
The resulting temperature profile for different choices of $\epsilon$ has been plotted in Fig. \ref{fig:wind_model}.

From Figure \ref{fig:wind_model}, it is clear that $\epsilon$ affects the heating pattern in three major ways: i) smaller $\epsilon$ (i.e. $\tau_{\rm{rad}}\ll \tau_{\rm{advec}})$ results in large temperature contrasts, whereas larger $\epsilon$ tends to horizontally homogenize temperature, increasing the night side temperature, ii) the temperature profile becomes more asymmetric about the maximum with increasing $\epsilon$, and iii) the temperature peak shifts in phase with increasing $\epsilon$. The phase shift of temperature maximum is to the left for $\epsilon>0$ and to the right for $\epsilon<0$.  To mimic the latitudinal variation in the wind, we posit a reversal at an angle $\theta_c$ to the companion equator. With $\epsilon > 0$ and large $\theta_c$, the model goes to a uniformly super-rotating companion wind.

\subsection{Characteristic Parameters for Companion Irradiation}
\label{sec:parameters}
Here we discuss physical parameters governing potential wind circulation in BWs and RBs. A main quantity of interest is the speed of heat redistribution for a given $\epsilon$, $v_{\rm w}= R_\ast \omega_{\rm adv}= R_\ast \epsilon /\tau_{\rm rad}$. We can estimate the radiative cooling time $\tau_{\rm{rad}}$ by
\begin{equation}\label{eq:tau_rad}
    \tau_{\rm{rad}} \equiv \frac{\rho c_PH}{\sigma T_0^3}~,
\end{equation}
with $\rho$ the density, $c_P$ the specific heat capacity and $H$ the thickness of the heated, advecting atmosphere layer. The radiative parameters are set by the properties of the companion heating zone.

The observed photon emission of the MSP is dominated by the LAT $\gamma$-ray flux, with a SED peaking at $\sim 3-10$\,GeV, representing a modest $\sim 1-30$\% fraction of the pulsar spindown power. We assume here that these photons dominate the companion heating, although the bulk of the pulsar emission is in the relativistic magnetized wind and, as noted above, the high energy particles from this wind (possibly reprocessed in an IBS) can also be important for the companion heating. For a direct heating model dominated by high energy $\gamma$-ray photons, we expect the energy $E_0$ to be deposited in an electromagnetic shower below the companion photosphere with energy distributed as
\begin{equation}\label{shower}
\frac{dE}{E_0 dz} = b (bz)^{(a-1)} e^{-bt}/\Gamma(a)~,
\end{equation}
with $a=1+b[{\rm ln}(E_0/Ec) + 0.5]$, $b\approx 0.5$, critical energy $E_c \approx 0.7{\rm GeV}/(1+Z)$ and the depth $z$ in units of the radiation length in matter of atomic number $Z$ and atomic mass $A$
\begin{equation}\label{radL}
    X_0 \approx 1430 {\rm g\,cm^{-2}}A/[z(Z+1)(11.3-{\rm ln}Z)]~. 
\end{equation} 
This means that for the MSP $\gamma$-rays, the energy is deposited over a depth $\Sigma \sim (10-12) X_0 \approx 700{\rm g\,cm^{-2}}$.

If we take column density $\rho H=\Sigma$ and use 
\begin{equation}\label{eq:cp}
    c_P\approx \frac{5}{2}\frac{k_B}{\mu m_p}= 3.5 \times 10^8\,\rm{erg\, K}^{-1}\,{\rm g}^{-1}~.
\end{equation}
For an ionized solar abundance plasma at temperature 5000$T_{5000}$K, we get an approximate heat flow speed of
\begin{equation}\label{eq:speed}
v_w\approx \frac{R_\ast \epsilon \sigma T_0^3}{\Sigma c_p} \approx 
2 \epsilon (R_\ast/0.1R_\odot) T_{5000}^3 \, {\rm km\, s^{-1}}~,
\end{equation}
or equivalently $\omega_{\rm rad} \approx (9.7\, {\rm h})^{-1}\epsilon T_{5000}^3$. This is the bulk motion averaged over the heated layer; the surface photospheric speed may be larger or smaller and in general may show a complex flow pattern. Since $C_s = 8 T_{5000}^{1/2}{\rm km\,s^{-1}}$ the bulk heat flow is subsonic at Mach number $M\sim 0.25 \epsilon T_{5000}^{5/2}$ except for very strong winds on very hot companions. Since we believe that in most cases BW and RB heating drives significant mass loss which forms the IBS shock, at some point above the photosphere,  there must be an escape velocity outflow with $v > 440 (M/R)^{1/2} \, {\rm km\, s^{-1}}$, with $M$ and $R$ in solar units.

Compared to hot Jupiters, a larger heating depth means that a large mass is involved in the flow and lower velocities can still effect very large heat re-distributions. The resulting surface brightness changes induce very substantial departures from the radial velocities expected from a uniformly direct heated companion (let alone from an unheated companion star where the center of light velocity $v_{\rm CoL}$ will be the same as the center of mass velocity $v_{\rm CoM}$). The $\sim {\rm km ~s^{-1}}$ shifts associated with the actual surface flow are more difficult to detect with current instrumentation.

\subsection{Numerical Implementation}
We have implemented this wind transport model in the {\tt ICARUS} code of \citet{breton2012koi}. In this code, the star surface is represented by a triangular tessellation. For every triangular tile, we collect all other tiles which fall within a latitude width of $\Delta \theta_0$. For each tile we assign E-W and W-E neighbors. To each tile, we then apply Eq. \eqref{eq:kinematic}, in discretised form
\begin{align}\label{eq: finitedifference}
    T_{i} = T_{i\pm 1} + \frac{1}{4T_{i\pm 1}^3}\left[1-\exp\left(-\frac{4T_{i\pm 1}^3\Delta\Phi_i}{|\epsilon|}\right)\right]\nonumber\\
    \times\left(f_{\rm{p},i} + T_N^4 - T_{i\pm 1}^4\right)~,
\end{align}
where $\Delta\Phi_i = |\Phi_i-\Phi_{i\pm 1}|$, and index $i+1$ and $i-1$ denote the E-W and W-E neighbour of $i$, respectively. The above equation reduces to the simple finite difference expression in the limit when $\Delta\Phi_i/|\epsilon|\ll 1$, but yields better convergence when the number of tiles in each latitudinal band is small (e.g. close to the poles of the star). Finally, in order to find the equilibrium temperature of all tiles, we repeatedly apply Eq. \eqref{eq: finitedifference}, each time with an updated temperature, until the system relaxes. This procedure is followed for each latitude $\theta$ on the companion.

In our modeling, we allow the night time temperature $T_N$ to be a free parameter; for RB this represents core nuclear luminosity, for BW tidal heating or deep circulation may provide the night side heat.

\begin{figure*}
    \centering
    \includegraphics[scale=0.50]{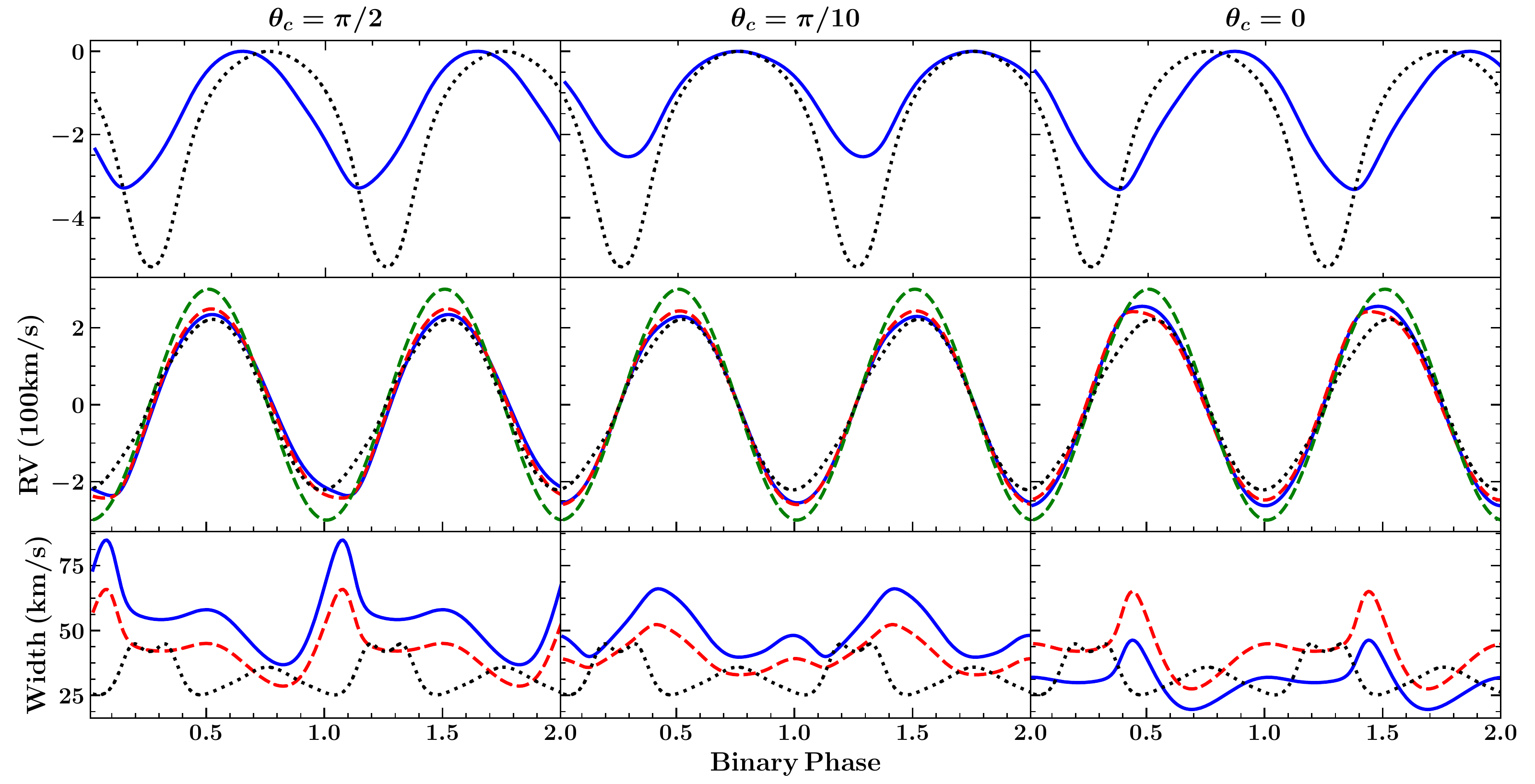}
    \caption{Wind heating effects for a binary system of $P_{\rm{B}}\sim 5$ hours, and $i\sim 60^{\circ}$. The left and right columns correspond to $|\epsilon| = 1$ super- and  sub-rotating winds, respectively. The central panels show $\theta_c$ adjusted so that, for the assumed $i=60^\circ$, flux from the equatorial band with the super-rotating winds nearly balances that from regions extending from mid-latitude to poles with the sub-rotating winds. Top: normalized light curves with (solid = WH) and without (dotted=DH) wind heat transfer. 
    Middle: Radial velocity curves for the companion center of mass (dashed green line), DH flux-averaged radial velocity (dotted black line), WH flux-averaged radial velocity with negligible surface motion (dashed red line) and WH including a global circulation flow of $+10$ km/s in the direction of the heat flow (solid blue line). Bottom: kinematic broadening (RMS line width) for the DH (dotted black), negligible velocity WH (dashed red) and +10km/s WH (solid blue).
    }
    \label{fig:lc_and_spectra}
\end{figure*}

\section{Light curve and Radial Velocity Effects}\label{obs_lc_spec}

From the relaxed companion temperature distribution, one can obtain light curves for a given viewing angle at binary inclination $i$. The detailed shape of the light curve will be highly sensitive to $\epsilon$ and $\theta_c$. For $\theta_c\rightarrow \pi/2$ and $\epsilon > 0$, the peak of the light curve will shift to earlier binary phase; for $\theta_c\rightarrow 0$, $\phi_{\rm max}$ increases. When $\theta_c$ takes on intermediate values, the contribution from opposing velocity flows at equatorial and mid-latitudes moves heat toward the night side at both the dawn and dusk limbs. The net effect can be a broader light curve with little or no peak shift.  This has been clearly shown in the top panels of Figure  \ref{fig:lc_and_spectra}.

Radial velocity measurements provide another probe of atmospheric circulation. Just as for light curves, heat redistribution directly affects the radial velocity curve for finite $\epsilon$. Recall that direct heating reduces the radial velocity amplitude $K$ by moving the center of light toward the nose of the companion. Figure \ref{fig:lc_and_spectra} shows the decrease from the center of mass $K_c$. Wind effects mitigate this, by moving heat away from the nose. In general, this increases the radial velocity amplitude from the DH case (but not to the CoM RV). However note that since the WH light curves are asymmetric, for some phases the flux weighted companion velocity competes with the orbital motion. This can lead to small regions of RV decrease (esp. $\epsilon >0$, $\phi_B\approx 0.8-0.9$). 

The ratio $\omega_{\rm{adv}}/\Omega$ of wind speed to the companion rotation speed is an important factor in determining spectral distortion from the wind compared to distortions created by simple companion co-rotation. Large $\omega_{\rm{adv}}/\Omega$ implies larger WH RV distortions. Since $\omega_{\rm{adv}}/\Omega\sim T_0^3/\Omega\sim L_{\rm{P}}^{3/4}a^{-3/2}\Omega^{-1}\sim L_{\rm{P}}^{3/4}$ (assuming that the star is tidally locked and $\Omega\sim a^{-3/2}$), the inferred spectral distortion for a fixed wind parameter $\epsilon$ should be largely independent of the binary period, but should increase with pulsar heating power. 

Observationally, the light curve distortions of the WH model translate directly to substantial non-sinusoidal terms in the radial velocity curves that can be measured in high-quality spectroscopic data. But any effect that redirects heating flux to these zones (e.g. magnetically ducted IBS particles \citep{2017ApJ...845...42S} or other star-spot inducing effects, see \citealt{2016ApJ...833L..12V}) will produce similar RV curve distortions. Accordingly, when showing application to two spiders in \S4 we compare WH effects with those from a simple hot spot (HS).

 Surface motions produce radial velocity shifts in addition to the illumination-induced effects. For small $v_W\ll K$ the changes are subtle and the non-sinusoidal pattern is similar to that induced by heating-induced temperature shifts. Interestingly, the effects on the kinematic line widths can be more striking, since, for $\epsilon \ge 0$, the surface velocity acts in opposition to the projected rotational velocity, resulting in broadened lines (here the RMS line width, equivalent to a Gaussian $\sigma$). For $\epsilon < 0$, the phase variations are smaller and the line widths are decreased. The bottom panel of Fig. \ref{fig:lc_and_spectra} shows how $\epsilon$ sensitivity produces phase variable line {\it widths}. Thus, high S/N high-resolution spectra may test the kinematic effects of WH with line-broadening studies. Note that we have computed simple flux-weighted broadening; additional effects from the lines' equivalent width temperature sensitivity can (modestly) modify these results. We discuss such effects below.
 
\begin{figure}
    \centering
    \includegraphics[scale=0.35]{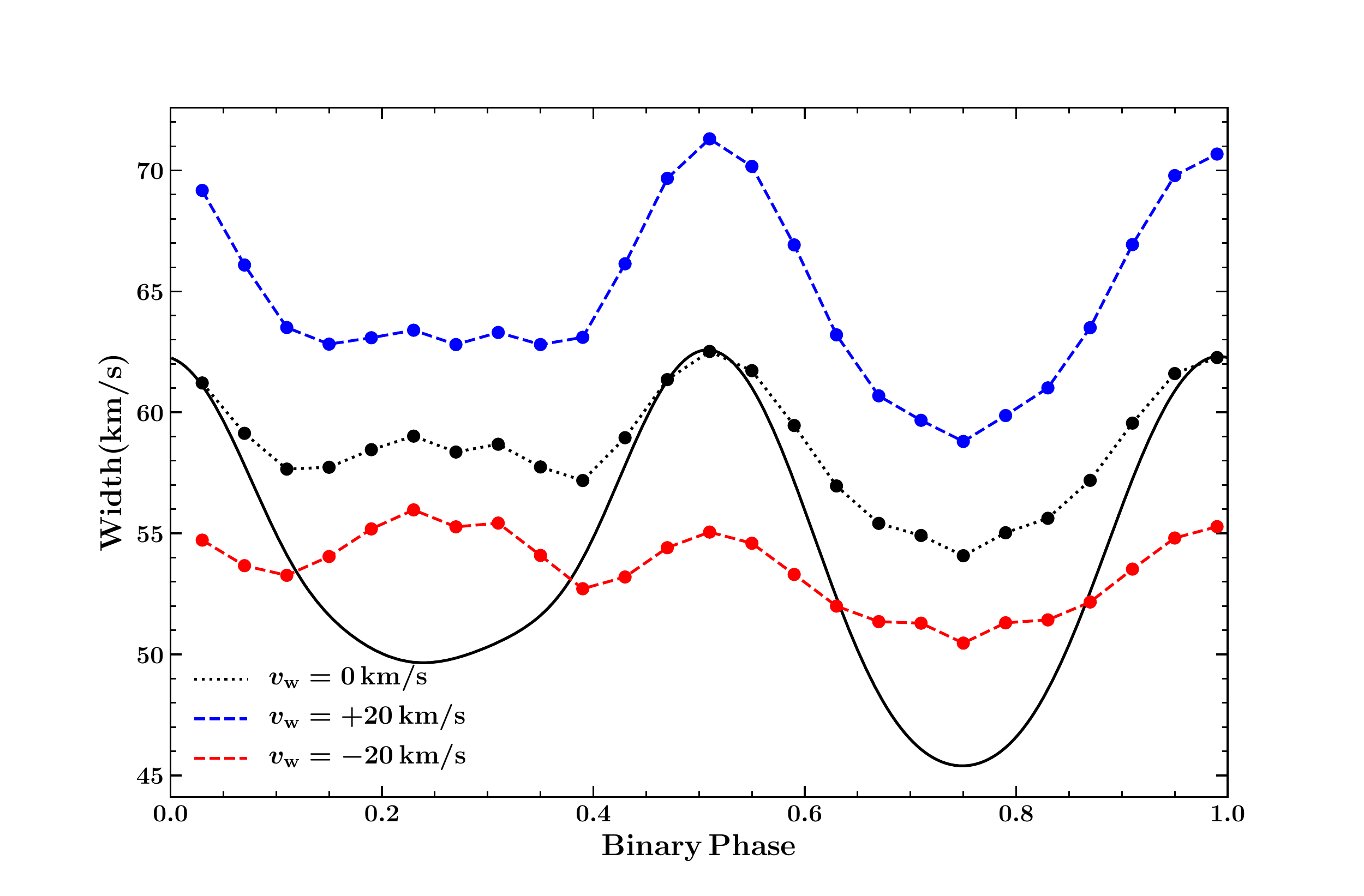}
    \caption{RMS line widths for finite observational time window, assuming $t_\mathrm{obs}/P_B = 0.04$, and a J2215-like $K_\mathrm{c} = 420$ km/s and $\epsilon=0.16$. The solid line assumes an infinitesimal observational window. The three curves show widths for three different cases: no surface motion (black), surface motion of +20 km/s (blue) and surface motion of -20 km/s (red).}
    \label{fig:width_smear}
\end{figure}

Finally, since these are relatively faint, short-period binaries, inevitably the spectroscopic exposure times $t_{\rm obs}$ will be a modest fraction of $P_B$. While distortions of the RV curve will be negligible for $t_{\rm obs}< P_B/20$, the effects on the velocity width are large. Figure \ref{fig:width_smear} illustrates this. Since the radial motion during the integration is largest when the surface velocity distribution is smallest, and vice-versa, integration time smearing decreases the variation $\sigma(\phi_B)$ unless the exposures are quite short. However, orbit-averaged surface velocity-induced width changes are relatively insensitive to $t_{\rm obs}/P_b$.

{}

\section{Applications of the Model}\label{application}
\subsection{PSR J1959+2048}
PSR J1959+2048 (hereafter J1959) is a BW  system with a $P=1.6$\,ms, $\dot{E}=  1.6\times 10^{35}\,{\rm erg\,s}^{-1}$ pulsar in a $P_b= 9.1$ hr orbit with a $\sim 0.03\,M_\odot$ companion. For the photometric fits, we use $BVRIK_s$ magnitudes from \cite{reynolds2007light} and an assumed extinction $A_V=0.82$, and for spectral analysis, we use radial velocity measurements published in \citet[][hereafter \citetalias{van2011evidence}]{van2011evidence}. This is the original BW pulsar and is of especial interest, since \citetalias{van2011evidence}, with an approximate treatment of DH effects, infer that it may have a pulsar mass as large as $2.4\pm0.12M_\odot$.

Table \ref{tab:fit_objects} shows the result of optical light curve fitting assuming three models: a standard direct heating model (DH), a model with atmospheric circulation (WH), and direct heating with a simple multiplicative hotspot (HS) which adjusts the underlying temperature by a factor $(1+A_{\rm HS})$ distributed as a Gaussian$^2$ over radius $r_{\rm HS}$, located at $(\theta_{\rm HS}, \phi_{\rm HS})$, with (0,0) the sub-pulsar point at the nose. When looking at the light curves, one can see that the phase shift in the maximum is small, although the peak is somewhat broadened and there is excess flux at phases $\phi\approx 0.75-0.9$, especially in the bluer bands. The WH model accommodates this with an intermediate $\theta_c\sim 20^\circ$, adjusting the maximum shape with little overall phase shift. The HS model in contrast adds blue flux in a narrow phase range by hiding a small, unrealistically hot ($5.3\times T$) spot in the Southern hemisphere, which pops out of eclipse to be visible only near $\phi \approx 0.9$. There appears to be no good HS solution with the spot in the Northern (visible) hemisphere. Thus while WH improves the fit by adjusting the LC over a broad phase range, the HS improvements come from a small region of phase. This is geometrically allowed but it seems more plausible to associate such narrow phase structure with an added non-thermal flux (or photometry errors). With fewer parameters WH is in any event statistically preferred: the DH model likelihood is just $0.5\%$ while the HS likelihood is $11\%$, according to the Akaike Information Criterion (AIC). Since the interpretation is also more physically plausible than a tuned eclipsing hot spot, we infer that WH is the best model for this data set. However it is important to note that, according to Table \ref{tab:fit_objects}, WH and HS have very similar values for the binary parameters. The principal difference is a somewhat lower base (night side) temperature for WH. Distance and therefore $L_p$ are somewhat lower for DH than the other models.

\begin{figure*}
    \centering
    \includegraphics[scale=0.7]{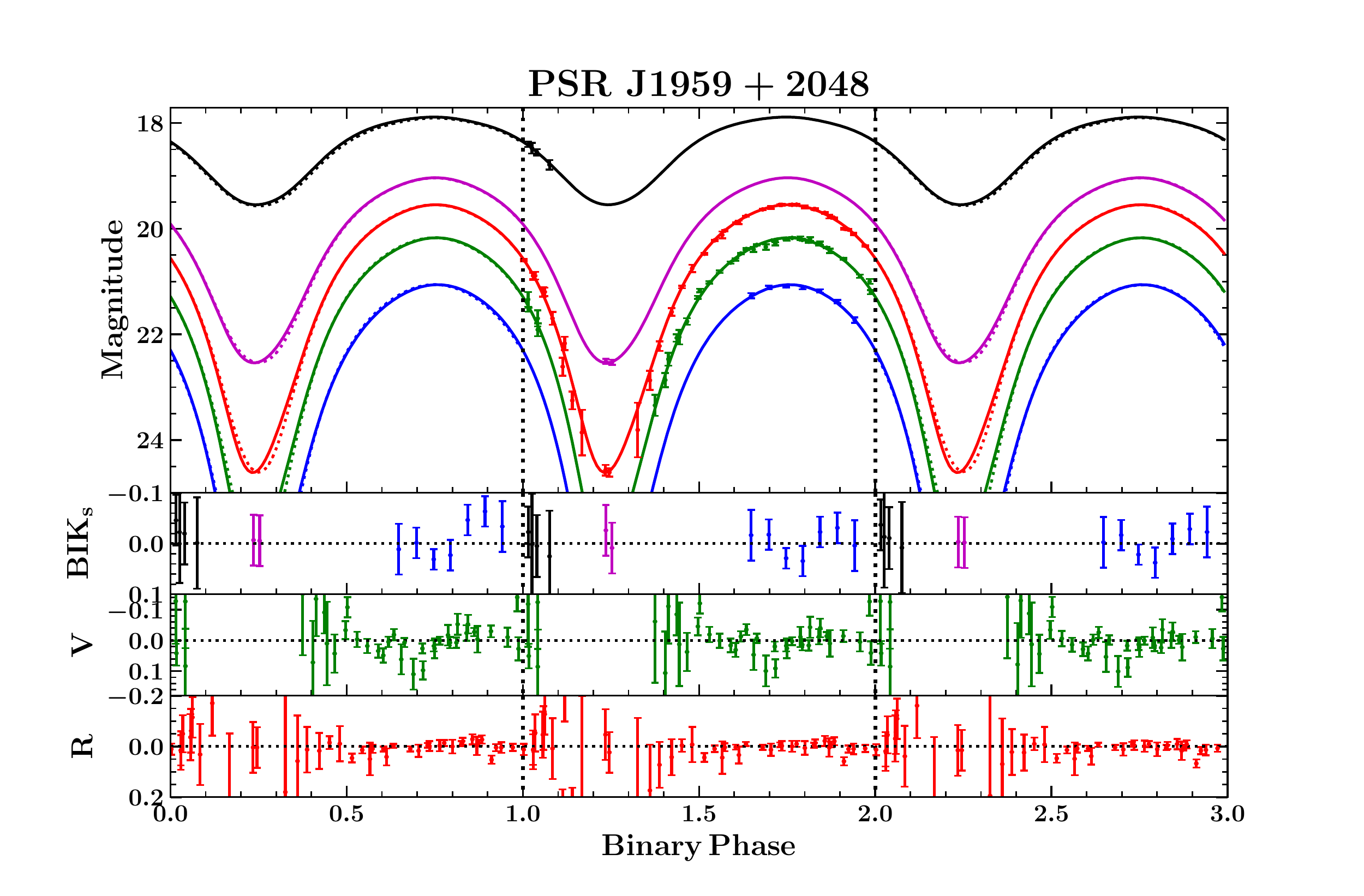}\hspace{0cm}
    \caption{{\it BVRIKs} light curves for J1959. Three periods are plotted with $\phi_B= 0$ at pulsar TASC (ascending node). Solid curves show wind heating (WH) model. The dotted curves of the first and the third cycles correspond to direct heating (DH) model and hotspot (HS) model, respectively. Lower panels show residuals from the DH model (first cycle), WH model (second cycle) and hotspot model (third cycle).}
    \label{fig:PSR_lc19}
\end{figure*}

The top panel of Fig \ref{fig:PSR_lc19} shows data (and other light curve models) compared with the best-fit WH model. The lower three panels show the fit residuals for the various models. An improved test of the model would require more data, especially for the brighter $I,\, Ks$ colors and especially towards the light curve minima, where the characteristic WH asymmetry is strongest. Hot spot models are best probed with bluer colors, so high quality $B$ or $u$ light curves covering maximum would also be useful. 

\begin{figure}
    \centering
    \includegraphics[scale=0.34]{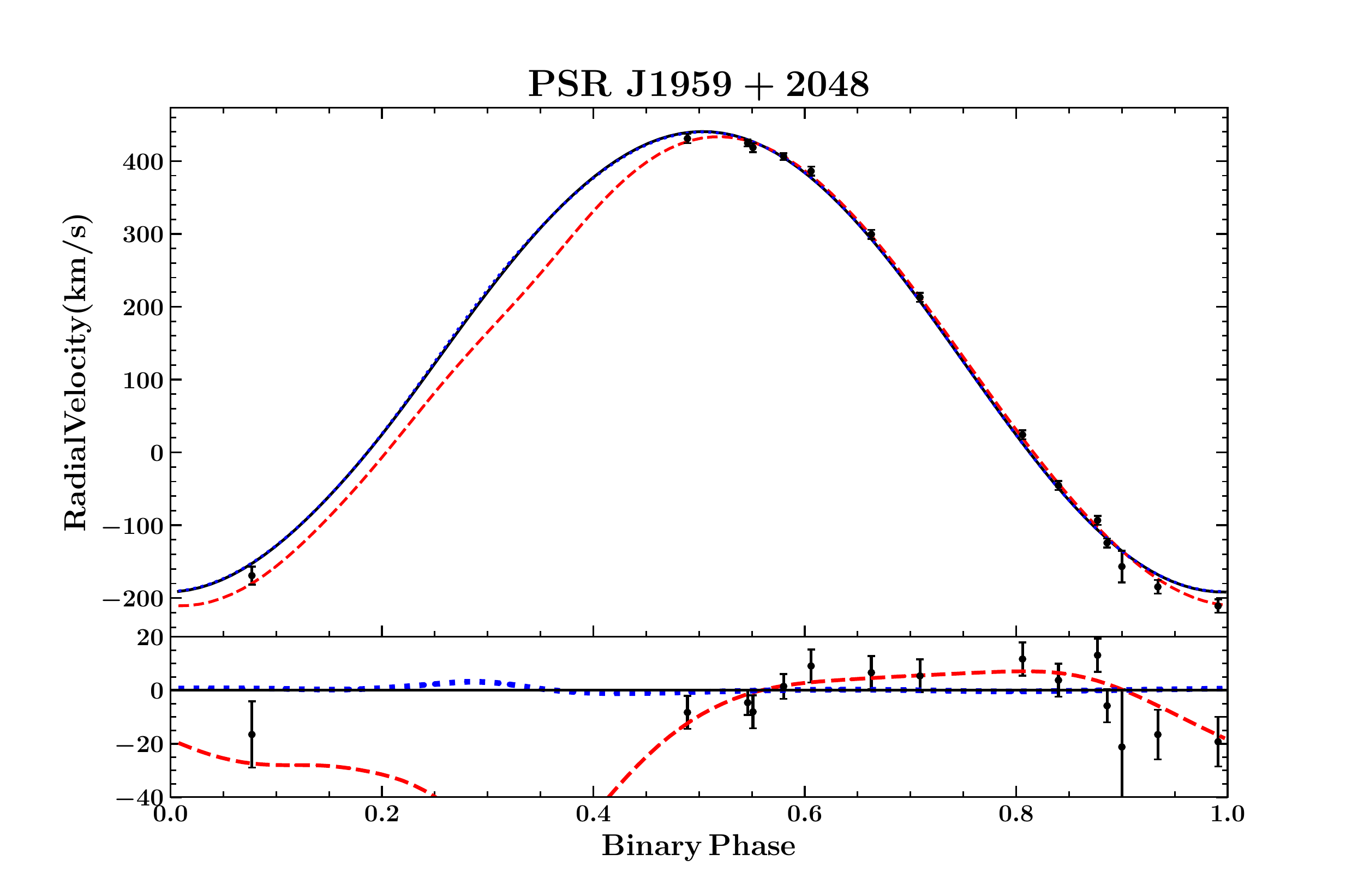}
    \caption{Radial velocity curve of PSR J1959. The black curve is obtained using best fit parameters assuming zero wind speed, the red dashed curve is by allowing variable wind speed, and the blue dotted curve is for the hotspot model. Lower panel: Residuals with respect to the best-fit zero speed wind model -- the points show the measurements, the red dashed line shows difference with the finite wind speed model curve, and the blue dotted curve shows the difference for a hotspot model.}
    \label{fig:PSR_sp19}
\end{figure}

We also analyzed the radial velocity data tabulated by \citet{van2011evidence}, which were measured against a G2 stellar template. The observed radial velocity amplitude $K_\mathrm{o}$ is sensitive to the heating pattern as well as surface speeds. As pointed out by \citetalias{linares2018peering}, absorption line equivalent widths depend on photospheric temperature (and density), and so different species are weighted by different $EW(T)$ across the face of the companion, giving different measured radial velocity amplitudes.  To estimate the true center of mass velocity $K_\mathrm{c}$, we fit the data assuming illumination determined by the several heating models. In the case of the WH model, we also consider a finite photosphere speed following the heat motion fit by the photospheric model fit. Each model generates $K_o$ curves, using the $EW(T)$ weighting (here using \citetalias{linares2018peering}'s MgI $EW(T)$ for the metal line-dominated template weighting, extrapolated to lower $T_{\rm eff}$ using archival template spectra, as needed) which are fit to the tabulated velocities. Although we do not perform a simultaneous photometry fit, we do marginalize the spectroscopic fit over the geometrical parameters from the end of the photometric MCMC chains, sampling $\sim 2\sigma$ uncertainties. Thus, the mass errors do include all uncertainties in the model fitting, spectroscopic and photometric. We have confirmed that these agree well with simple estimates for the masses produced by the propagation of the individual parameter errors. The results for the fitting are shown in Table \ref{tab:fit_objects}, and the radial velocity curves along with their residuals for different models are presented in Fig. \ref{fig:PSR_sp19}. With our base model of WH with $v_w=0$ (and MgI EW weighting), the best fit $K_c$ is $\sim 338$ km/s, slightly smaller than $\sim 340$ km/s predicted with the simple phase-shift model. This small decrease in $K_c$ for our base model together with a small increase in $i$ (see Table \ref{tab:fit_objects}) brings our estimated mass down to $2.18\, M_\odot$ from $2.29\, M_\odot$. Interestingly, if we allow the surface speed to be a free parameter in our model, its best fit value is $\sim 33$ km/s, which is $\sim 4\times$ larger than the sound speed, and $>40\times$ the estimate of Eq. \eqref{eq:speed}. We also check the statistical likelihood of the three models. According to the AIC, the DH, HS and WH model (with no surface speed) are not distinguished. WH with the large 33 km/s surface flow is preferred (zero velocity models have an AIC relative likelihood of 0.04\%). The result of this large surface speed is a decrease in $K_c$ to $\sim 326$ km/s, and therefore a decrease in the estimated neutron star mass to $\sim 1.95\, M_\odot$. However, given that the night phase spectroscopy is very limited, that Mach$\sim 4$ flow would be quite puzzling and that the preference for finite velocity is not highly statistically significant, we still consider zero speed WH to be our preferred model; additional spectroscopy is needed before large surface speeds should be invoked.  

\begin{deluxetable}{lccc}
\tabletypesize{\footnotesize}
\tablewidth{0pt}
 \tablecaption{Light Curve fit results for J1959\label{tab:fit_objects}}
 \tablehead{
 \colhead{Parameters} & \colhead{DH} & \colhead{WH} & \colhead{HS}}
 \startdata 
 $i (\mathrm{ deg})$ & $62.5^{+1.3}_{-1.1}$  & $63.8^{+1.3}_{-1.2}$ & $63.6^{+1.3}_{-1.1}$\\
  $f_c$ & $0.900^{+0.010}_{-0.010}$  & $0.898^{+0.008}_{-0.007}$ & $0.894^{+0.009}_{-0.009}$ \\
 $L_{\mathrm{P}}/10^{34}(\mathrm{erg}/\mathrm{s)}$ & $3.01^{+0.04}_{-0.04}$  & $4.82^{+0.12}_{-0.11}$ & $4.74^{+0.10}_{-0.10}$ \\
 $T_N$ (K) & $2670^{+30}_{-30}$ &$2488^{+78}_{-86}$ & $2734^{+31}_{-29}$\\
 $d_{kpc}$ & $2.04^{+0.01}_{-0.01}$  & $2.22^{+0.03}_{-0.02}$ & $2.27^{+0.02}_{-0.02}$\\
 $\epsilon$ & -& $-0.25^{+0.03}_{-0.03}$ & - \\
 $\theta_c (\mathrm{deg})$ &-&$19.2^{+0.5}_{-0.7}$ & -\\
 $\theta_{\rm{hs}}\,(\rm{deg})$&-&-&$57.0^{+5.7}_{-6.5}$\\
 $\phi_{\rm{hs}}\,(\rm{deg})$&-&-&$-71.7^{+7.9}_{-5.4}$\\
 $\mathcal{A}_{\rm{hs}}$&-&-&$4.3^{+2.4}_{-2.3}$\\
 $r_{\rm{hs}}\,(\rm{deg})$&-&-&$7.5^{+4.6}_{-3.1}$\\
 $\chi^2/\mathrm{DoF}$ & $140/89$  & $119/87$ & $118/85$
 \enddata
\end{deluxetable}

\subsection{PSR J2215+5135}

PSR J2215+5135 is a $P_s = 2.61$\,ms, ${\dot E} = 5 \times 10^{34} {\rm erg\, s^{-1}}$ `redback' MSP in a 4.14\,h orbit about a $\sim 0.3\,M_\odot$ companion. Optical photometry shows a large day-night flux difference, indicating that the companion star is strongly irradiated by the pulsar. Moreover, there is evidence of an orbital phase-shift in the light curve maximum of this system (see \citealt{schroeder2014observations, romani2015keck}, hereafter \citetalias{romani2015keck}), so heating asymmetries are strongly indicated. The system parameters are also of interest since \citet[][hereafter \citetalias{linares2018peering}]{linares2018peering}
infer $M_{\rm NS}=2.27^{+0.17}_{-0.15}M_\odot$, although \citetalias{romani2015keck} infered a much larger inclination and, hence a smaller $M_{NS} = 1.6 M_\odot$.

LSC18 present William Herschel Telescope (WHT) ACAM $g^\prime r^\prime i^\prime$ measurements from two nights, along with some IAC-80 $g^\prime r^\prime$ magnitudes. To ensure that the photometry is on a consistent flux and phase scale, we downloaded the WHT J2215 images and associated calibration frames from the archive, calibrated the images and performed photometry on the pulsar and a grid of nearby field stars. These field stars have PanSTARRS catalog fluxes, so we converted these to SDSS $g^\prime r^\prime i^\prime$ magnitudes and used these to calibrate our pulsar photometry. The exposure midpoint times were barycentered and phased with a LAT ephemeris that allows an accurate pulse fold from 2008.7 to 2017.1 (L. Nieder, private communication). The resulting light curves are consistent with those of \citetalias{linares2018peering}, but have estimated photometric systematic errors smaller than $\delta m \approx 0.03\,$mag and negligible phase uncertainty.  We were not able to obtain the IAC-80 frames. These are interesting because they fill in gaps in the WHT light curves, but all we can do is use the photometry and estimated arrival phases plotted by \citetalias{linares2018peering}. We find that substantial zero point shifts are needed ($\delta g^\prime = 0.029$\,mag and $\delta r^\prime = 0.034$\,mag) to match the IAC magnitudes to the WHT data at the phase overlaps. This may be attributed to fluxing errors, but might also be due to intrinsic variability between epochs. Indeed the WHT data also show evidence for variation between the two nights (see below). Accordingly, we use only the WHT data in the fits, but plot the IAC points for comparison. We also found some PanSTARRS DR2 individual detections of the pulsar (PS2: 14 in $g^\prime$, 7 in $r^\prime$ and 12 in $i^\prime$) so, for completeness, we converted these to SDSS magnitudes and phased the resulting points. The results were broadly consistent with the expected magnitudes, but had much larger errors and, since they were obtained over several years, may well include substantial source variability. We thus ignore these in further analysis. 
 
\begin{deluxetable}{lccc}
\tabletypesize{\footnotesize}
\tablewidth{0pt}

 \tablecaption{Light Curve fit results for J2215\label{tab:fit_J2215}}

 \tablehead{
 \colhead{Parameters} & \colhead{Phase-shift} & \colhead{WH} &\colhead{HS}}
 \startdata 
 $i (\mathrm{ deg})$ & $58.1^{+1.9}_{-1.9}$  & $68.3^{+1.8}_{-1.7}$ & $68.9^{+1.9}_{-1.8}$ \\
  $f_c$ & $0.92^{+0.01}_{-0.01}$  & $0.93^{+0.01}_{-0.01}$ & $0.94^{+0.01}_{-0.01}$\\
 $L_{\mathrm{P}}/10^{34}(\mathrm{erg}/\mathrm{s)}$ & $5.4^{+0.3}_{-0.3}$  & $4.2^{+0.2}_{-0.1}$ & $3.9^{+0.2}_{-0.1}$\\
 $T_N$ (K) & $5623^{+22}_{-24}$ &$5684^{+14}_{-16}$ & $5682^{+14}_{-15}$\\
 $d_{kpc}$ & $3.61^{+0.07}_{-0.07}$  & $3.30^{+0.04}_{-0.04}$ & $3.30^{+0.04}_{-0.04}$ \\
 $\epsilon$ & -& $0.16^{+0.08}_{-0.04}$&-\\
 $\theta_c (\mathrm{deg})$ &-&$35.0^{+19.3}_{-7.2}$&-\\
 $\Delta\phi$ & $-0.008^{+0.001}_{-0.001}$ & -&-\\
 $f_2 (\mu\mathrm{Jy})$ & $1.1^{+0.2}_{-0.2}$ & $1.2^{+0.2}_{-0.2}$& $1.2^{+0.2}_{-0.2}$\\
 $\theta_{\rm{hs}}\,(\rm{deg})$&-&-&$324.0^{+10.1}_{-8.7}$\\
 $\phi_{\rm{hs}}\,(\rm{deg})$&-&-&$73.2^{+5.0}_{-8.4}$\\
 $\mathcal{A}_{\rm{hs}}$&-&-&$0.6^{+0.4}_{-0.2}$\\
 $r_{\rm{hs}}\,(\rm{deg})$&-&-&$15.5^{+3.9}_{-6.8}$\\
 $\chi^2/\mathrm{DoF}$ & $489/235$  & $336/234$ & $297/232$
 \enddata
\end{deluxetable}

\begin{figure*}
    \centering
    \includegraphics[scale=0.7]{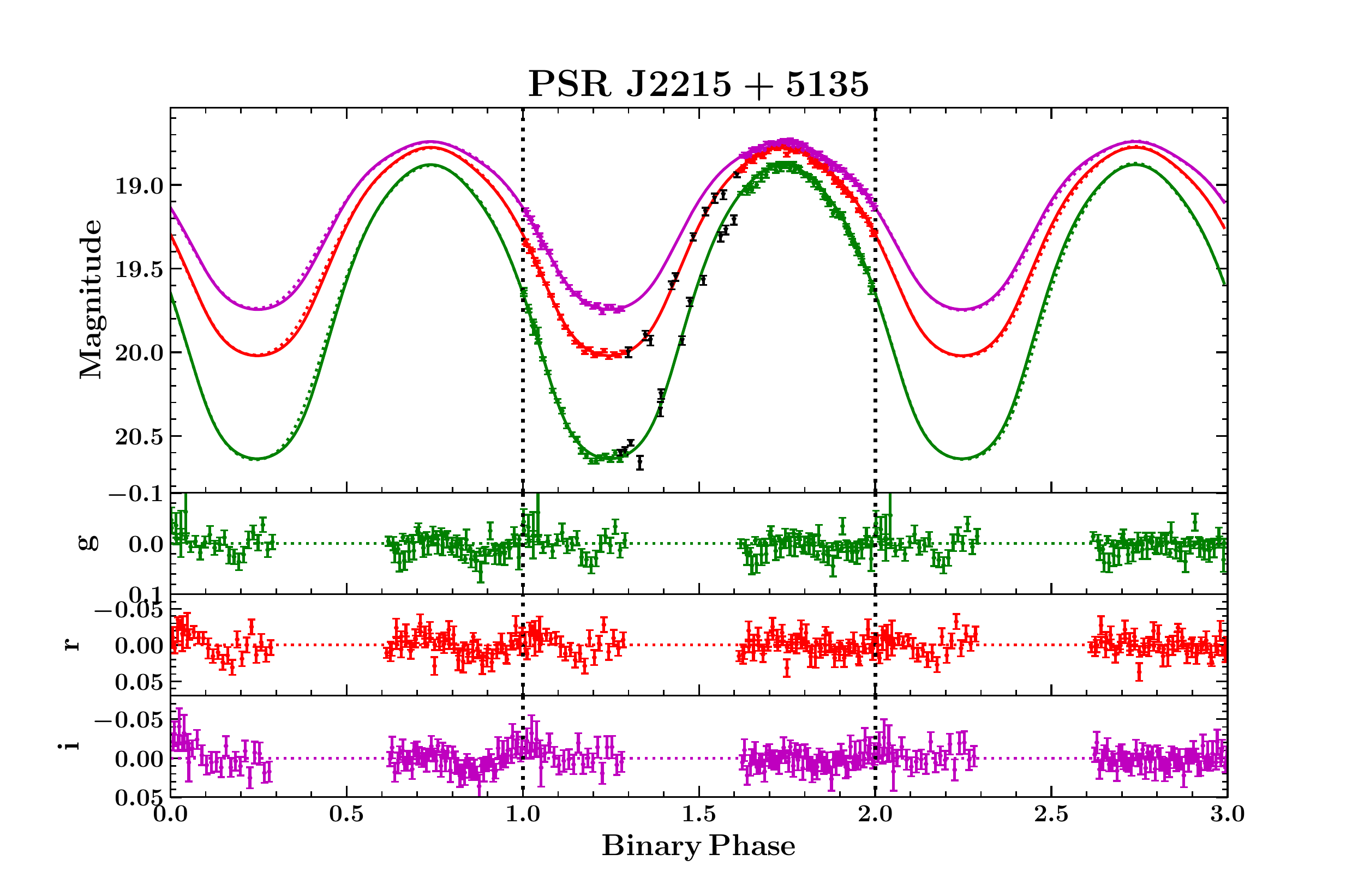}\hspace{0cm}
    \caption{{\it g'r'i'} light curves for J2215, with colored points for the WHT data. IAC-80 points are shown in black, but were not used for model fitting. Three periods are plotted with $\phi_B= 0$ at pulsar TASC (ascending node). Solid curves show wind heating (WH) model. The dotted curves of the first and the third cycles correspond to DH model with the arbitrary phase shift (a fit without phase shift is completely unacceptable) and HS model, respectively. Lower panels show fit residuals from the DH (w/ shift), WH and HS model.}
    \label{fig:PSR_lc22}
\end{figure*}

\begin{figure}
    \centering
    \includegraphics[scale=0.34]{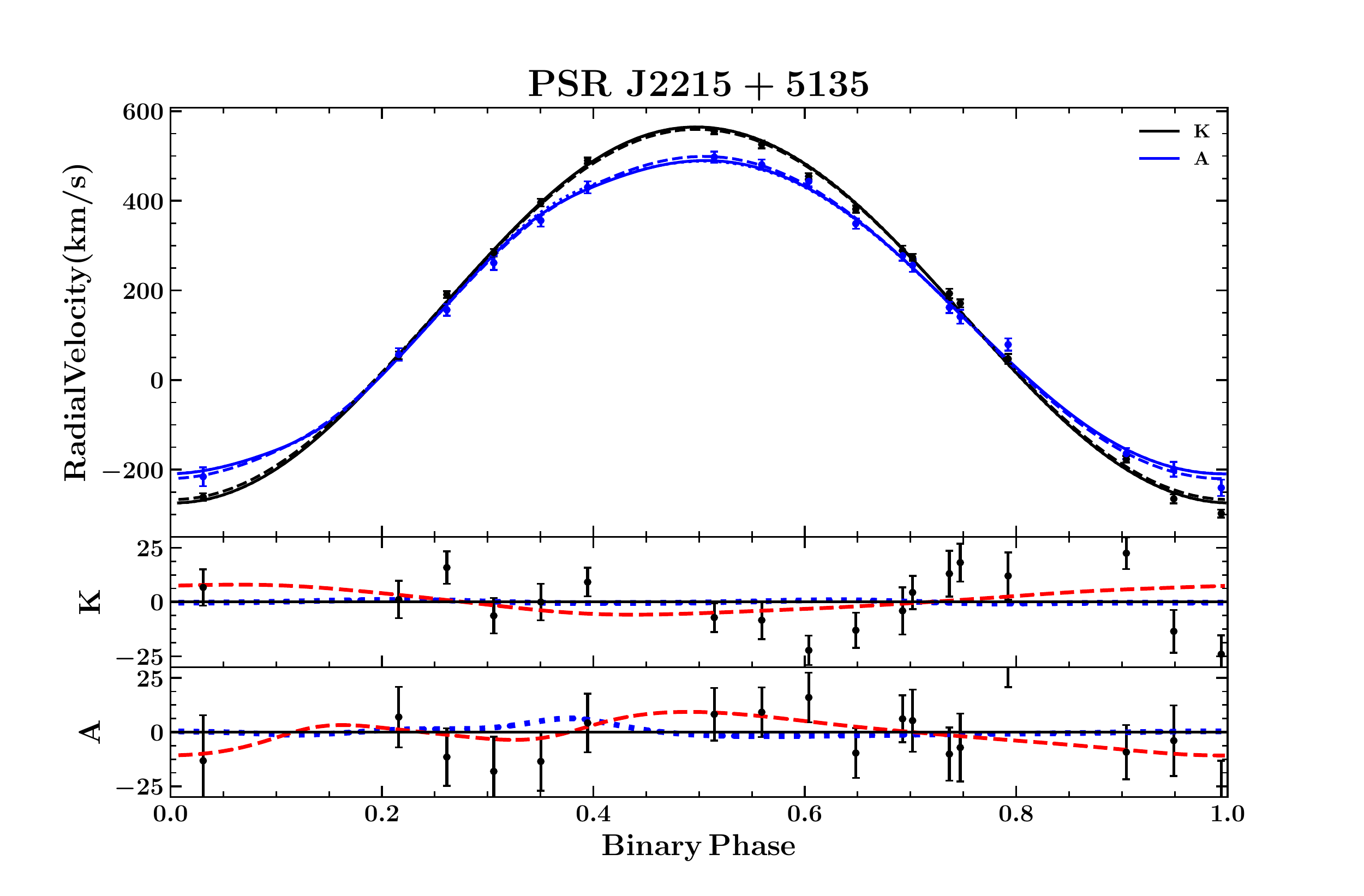}
    \caption{Upper panel: Radial velocity curve of PSR J2215 for cross-correlation measurements with K star and Balmer-line (A star) templates. The solid curves show the best-fit wind model with zero surface speed, the dashed curve shows the best wind model with finite surface speed and the dotted curve shows the hotspot model. Bottom panels: Residuals with respect to the best-fit zero speed wind model -- the points show the measurements, the red dashed line shows difference with the finite wind speed model curve, and the blue dotted curve shows the difference for a hotspot model.}
    \label{fig:PSR_sp22}
\end{figure}

\begin{deluxetable*}{lcccccccc}
\centering
\tabletypesize{\footnotesize}
\tablewidth{0pt}
 \tablecaption{Radial velocity fit results\label{tab:fit_objects}}
 \tablehead{
 \colhead{Parameters} &\colhead{J1959 DH/$\phi_B$-shift} & \colhead{Wind} & \colhead{Wind ($v_w$)}&  \colhead{HS}& \colhead{J2215 DH/$\phi_B$-shift} & \colhead{Wind} & \colhead{Wind ($v_w$)} & \colhead{HS}}
 \startdata 
 $K_\mathrm{c}\,(\mathrm{km/s})$ & $339.5^{+3.6}_{-3.7}$ & $338.3^{+3.6}_{-3.6}$ & $325.6^{+4.0}_{-4.0}$ & $337.6^{+3.7}_{-3.7}$ & $431.5^{+3.9}_{-3.8}$ & $429.4^{+3.9}_{-3.8}$& $421.8^{+5.3}_{-5.4}$ & $429.8^{+3.9}_{-3.9}$\\
 $\Gamma\,(\mathrm{km/s})$ & $124.4^{+3.0}_{-2.9}$& $123.9^{+2.6}_{-2.6}$& $120.5^{+1.9}_{-1.9}$& $124.4^{+2.7}_{-2.7}$& $142.8^{+3.3}_{-3.3}$ & $142.4^{+3.3}_{-3.3}$&$142.3^{+3.3}_{-3.4}$&$142.4^{+3.3}_{-3.3}$\\
 $v_\mathrm{w}\,(\mathrm{km/s})$ &-&-&$33.6^{+9.0}_{-8.5}$&-& -&-&$-30.6^{+14.9}_{-16.1}$&-\\
 $M_{\mathrm{NS}}\,(M_\odot)$& $2.29^{+0.10}_{-0.10}$& $2.18^{+0.09}_{-0.09}$& $1.95^{+0.09}_{-0.08}$ & $2.18^{+0.09}_{-0.09}$& $3.05^{+0.17}_{-0.18}$& $2.28^{+0.10}_{-0.09}$& $2.16^{+0.11}_{-0.10}$&$2.24^{+0.09}_{-0.09}$\\
 $M_{\mathrm{c}}\,(M_\odot)$& $0.034^{+0.001}_{-0.001}$&$0.033_{+0.001}^{-0.001}$ & $0.030^{+0.001}_{-0.001}$ &$0.033_{+0.001}^{-0.001}$ & $0.42^{+0.02}_{-0.02}$ &$0.31^{+0.01}_{-0.01}$ & $0.30^{+0.01}_{-0.01}$ &$0.31^{+0.01}_{-0.01}$ \\
 $\chi^2/\mathrm{DoF}$&28/12& 28/12& 8/11 &29/12& 68/33& 67/33& 62/32 & 72/33
 \enddata
\end{deluxetable*}

\citetalias{linares2018peering} find that the light curve fit is significantly improved by allowing an extra flat spectrum $\sim 35-70\mu$Jy, phase-independent flux to be added to each filter. We do not find such a large added flux, but we do notice a significant increase in the flux of the second WHT night (September 01, 2014), which we cannot attribute to a calibration error. We can fit this as a flat spectrum $f_2\sim (1.2\pm 0.2)\mu$Jy addition. The overall model also prefers a $\Delta m_r\sim 0.031\pm 0.001$\,mag offset in the night two magnitudes and so we include this as a free parameter in the fits, covering a possible photometric zero-point error. The night one WHT data are consistent with no veiling flux.

In fitting these data, we fixed $A_V$ to 0.40, as estimated from 3D dust maps \citep{2018MNRAS.478..651G}. If fitted as a free parameter, we find $A_V\sim 0.42\pm 0.03$. As noted above we allow a veiling flux (and $r'$ calibration offset) for night two. Table \ref{tab:fit_J2215} shows the best fit parameter results. We start with simple direct heating (DH) model. This model is completely unacceptable unless we allow an arbitrary phase shift $\Delta \phi$ (as an additional free parameter). The wind model parameters give a super-rotating equatorial wind and sub-rotating mid-latitude wind which result in the overall phase-shift of the light curve maxima by $\Delta\Phi_{\rm{max}}\sim -0.009$. While the $\chi^2$ decrease of this model is large, an even better fit is obtained with a simple hot spot model, invoking a plausible 60\% temperature increase in a $15^\circ$ radius spot to effect a similar phase shift. Note that the best-fit second-night veiling flux and $r$ zero-point shift are the same for all the models; our results for $i$ and $T_{\rm N}$ are quite independent of the presence or absence of this small veiling flux, although the fit quality improves significantly when these parameters are included. AIC prefers HS over WH at the $10^{-5}$ level, despite the higher number of parameters (this is possible, because of the many degrees of freedom). We conclude that for this RB a hot spot (likely associated with a magnetic pole) model is preferred.

Fig. \ref{fig:PSR_lc22} shows the light curve for the best fit WH model as a solid line and dotted lines for the DH+phase shift and HS models in panels 1 and 3 respectively. Fit residuals for the three models are shown in the lower panels. The upper panels show that the largest model differences happen to be at $\phi = 0.3-0.6$, where we have only the less reliable IAC-80 data; clearly precision photometry in this range would allow even better model discrimination. Nevertheless, we can see how the WH/HS patterns absorb most of the light curve structure, while providing a physical origin for the large phase shift. Of course, even HS is not a perfect match, showing e.g. small $g$ deficit near $\phi=0.2$. Additional emission components, such as other small hot spots or non-thermal IBS emission would be needed for a perfect fit. But with $\chi^2/DoF = 1.28$ for HS and nearly identical binary parameters for WH and HS, it is likely that these fits already robustly describe the heating and binary viewing geometry. 

For the spectroscopic analysis, we re-measured the DEIMOS/LRIS Keck spectra of \citetalias{romani2015keck}. For comparison, we also examined the radial velocities obtained from Balmer and MgI-triplet absorption lines reported by \citetalias{linares2018peering}. Although we did not re-measure their spectra, we used the archive exposure times and our ephemeris to recompute the phases of the exposure midpoints, with negligible error. The phases are quite close to those of \citetalias{linares2018peering}, although we do note that they omitted measurements of one of their spectra (at $\phi_B=0.199$).

HI and MgI absorption line strengths are weighted differently across the face of the companion, giving different observed $K_o$ radial velocity amplitudes for the same $K_c$. Since J2215 displays a large range of $T_{\rm eff}$ through the orbit, this can be a substantial effect. To follow such effects, we extracted radial velocities cross-correlating against G0 and K0 templates, where the cross correlation is dominated by neutral metal lines, such as the MgI triplet. To follow the Balmer velocities, we cross correlated with an A2 template (comparable to the temperature at maximum), restricting to $\pm 100$\AA\, around H$\alpha\beta\gamma\delta$.

We thus model the observed velocities with model $K_o$ computed for a given $K_c$, by averaging up the radial velocities over the companion surface elements, weighted by the received $g+r$ flux (our dominant spectral range) times the temperature-dependent $EW(T)$ of LSC18 for Balmer and MgI lines, as appropriate, and fit to the cross-correlation velocities. As for J1959, the fits are marginalized over the photometric parameter uncertainties. We show the radial velocity curves along with their residuals for different models in Fig. \ref{fig:PSR_sp22}, and also report the results as the corresponding Center-of-Mass radial velocity $K_\mathrm{c}$, $v_w$ and systemic velocity $\Gamma_\mathrm{c}$ (see Table \ref{tab:fit_objects}). Our templates were not well calibrated to the local standard of rest, so the $\Gamma_\mathrm{c}$ are not meaningful -- indeed we find small shifts between the different templates. 

Our base wind fit assumes negligible surface velocity $v_w$, but we can also allow a finite surface speed. Here `+' indicates global circulation in the direction of the local heating flow (e.g. positive at equatorial latitudes and negative at mid-latitudes, for super-rotation). In this case, we obtain the best fit $v_w\sim -30\pm 15$ km/s. This is five times larger in magnitude than estimated in Equation \eqref{eq:speed}, with a direction opposite to the heat flow. We also check the statistical likelihood of the three models. According to the AIC, the DH model and the WH model (with no surface speed) have likelihoods of 62\% and 81\%, respectively, compared to the WH model with surface speed. Thus for the RV data alone no model is strongly preferred.

\begin{figure}
    \centering
    \includegraphics[scale=0.28]{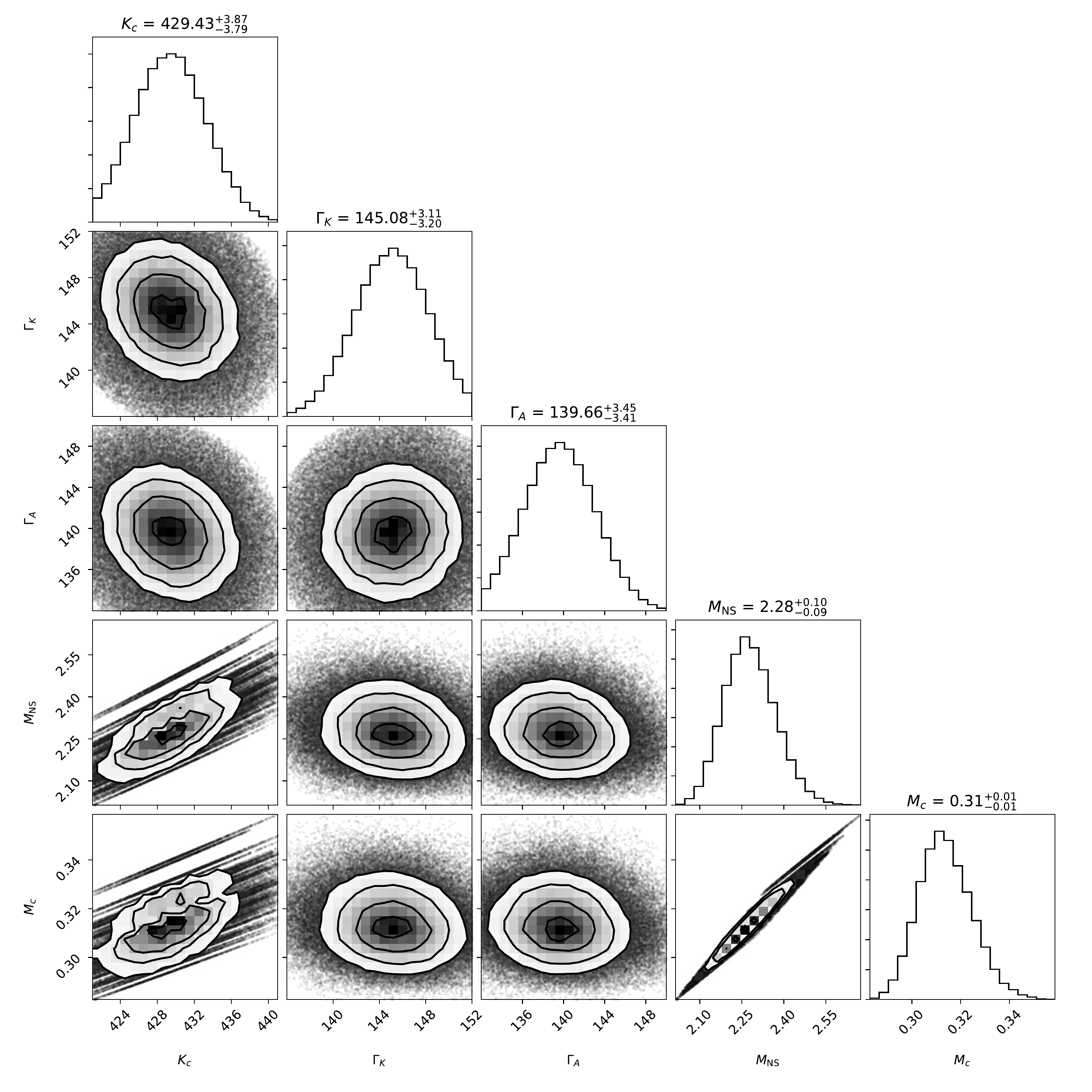}
    \caption{Corner plot showing distribution of parameters from our RV fitting for J2215 assuming a wind model with negligible surface speed. $\Gamma_K$ and $\Gamma_A$ are the K0 (Mg I) and A2 (Balmer) template systemic velocities.}
    \label{fig:corner_plot}
\end{figure}

In view of the results in Figures \ref{fig:lc_and_spectra} and \ref{fig:width_smear}, we would like to better constrain $v_w$ by using the companion line widths. Unfortunately, our J2215 spectra lack the S/N to allow detailed profile fitting of the strong Balmer lines, even at maximum, and the individual weak metal lines are much too noisy for such analysis. We attempted to extract mean line widths from the cross-correlation analysis. To do this, we applied $\sigma=0-100$km/s artificial broadening to the spectra of the  G-type comparison star, simultaneously observed through the slit, added Poisson noise to return the convolved spectrum to its original signal to noise level, then applied cross-correlation analysis to these spectra. This gives the spectrum-by-spectrum dependence of the cross-correlation width on the companion velocity dispersion, Unfortunately, with 0.62\AA/pixel (Resolution $\approx 89$km/s) for our Keck spectra, we lacked the resolution to probe the expected $\sigma = 50-60$\,km/s line widths. In the best set of Keck LRIS data covering $\phi_B=0.51-0.74$, we found a mean width of $\sigma = 74 \pm 5 ({\rm scatter})\pm 4 ({\rm calib.\, variability})$km/s. But since the cross-correlation peak width showed little to no sensitivity below $\sigma= 50$\,km/s, we conclude that the evidence for larger companion velocity widths is weak, at best. Higher spectral resolution data, with good S/N will be needed to make the required measurements.

\subsection{Comparison with Previous Fits}

For J1959, our direct heating fits are quite similar to those of \citetalias{van2011evidence}. Our best fit $i$ is slightly smaller than their $i=(65\pm2)^\circ$ estimate, but our smaller CoM correction (assuming MgI-like equivalent width weighting) more than compensates, leaving a smaller mass. The superior WH fit implies a slightly larger $i$ and a further $1\sigma$ decrease in the best-fit pulsar mass to $2.18\pm 0.09 M_\odot$: nearly identical values are found for the best-fit HS model. If we allow large surface velocity in the WH model, the mass can drop below $2.0M_\odot$, but this requires a highly supersonic photosphere speed, larger than the mean speed of the heating zone estimated in Eq. \eqref{eq:speed}. Supersonic motions are modeled for hot Jupiter winds (see \citealt{fromang2016shear}), so flows with Mach numbers of a few may be possible on strongly heated spider companions. It is less clear that the surface photospheric speeds can exceed the speed of bulk heat transport in the absorption zone (Eq. \ref{eq:speed}). However one can imagine a Hadley cell-like circulation with larger, mildly supersonic speeds in the photospheric layers averaging to slower heat transport via return flow in deeper layers. In principle, one might have a surface overshoot giving a negative photospheric velocity (as fit for J2215), but this seems rather contrived. 

Improved photometry can distinguish the WH and HS heating patterns. High precision photometry near orbital minimum can detect the characteristic asymmetry imposed by wind flow, while more detail measurements of blue colors near maximum are sensitive to hot spots. For example, the small $M_{\rm NS} \approx 1.7M_\odot$ allowed by the magnetically ducted heating model of \cite{2017ApJ...845...42S} resulted from a very small particle-heated region needed to allow $d$ well under 2kpc. High precision $B$ or $u$ photometry should reveal such structure, which is not indicated in the present analysis. Most important, though, are improved spectroscopic constraints on WH surface motions. Figure \ref{fig:PSR_sp19} suggests that RV measurements during the night half of the orbit, where departures from a sinusoid are large, could help. Alternatively, high-resolution measurements of line shape near mid-day could be useful. 

For J2215, our fit of $i= 58.1^\circ$ for the simple phase-shift model differs slightly from the equivalent value of $i\sim 63.9^\circ$ in \citetalias{linares2018peering}. Since these are both measured principally from the WHT photometry and because of the potential importance of the source for neutron star mass measurements, we discuss the origin of this discrepancy in the appendix. In \citetalias{romani2015keck}, a much larger $i$ was found. We refit the 2010-2011 \citet{schroeder2014observations} MDM photometry analyzed in that paper with the DH model and confirm that this fit gives $i \approx 88^\circ$ and $T_N \approx 6300$\,K; howver, as noted by LSC18, this $T_N$ is inconsistent with the night-side spectra. This was evidently due to a large veiling flux at that epoch. Including such a flat spectrum component, which the model fits to $11.9\pm0.1 \mu$Jy, the inclination drops to $i=62.4\pm 2.4^\circ$ and $T_N$ drops to $(5850\pm 100)$\,K. The fit improves by $\Delta \chi^2=-15$. The large 2010-2011 veiling flux can, therefore, explain the fit values (and small neutron star mass) inferred in \citetalias{romani2015keck}. In contrast, our 2015 Keck spectra set a limit of $\sim 1.5\mu$Jy on the line veiling flux at Ca H\,\&\,K at that epoch. 

We conclude that in spider binaries there can be an important, and variable, veiling flux. When bright, this can substantially affect the light curve minima. With a blue non-thermal spectrum this might be plausibly associated with the IBS. Detailed studies of spider light curves should allow for such veiling. Since IBS emission might not be phase independent, it may be challenging to isolate this emission from light curves alone. But veiling may certainly be revealed by high-quality phase-resolved spectra. For example, our Keck data show that at our spectroscopic epoch we did not experience the very large ($35-70\mu$Jy) veiling flux quoted by \citetalias{linares2018peering}.

We find that our {\it observed} J2215 radial velocity amplitudes are quite similar to those of \citetalias{linares2018peering}, for both neutral metal lines and Balmer-dominated estimates. However, the model corrections to $K_c$ are quite different and we discuss this discrepancy in the appendix. We infer that, with a DH model one should derive an unphysically large $\sim 3 M_\odot$ mass for the pulsar and that LSC18's treatment may be in error. However, the WH model fit gives an interesting, but plausible $2.28^{+0.10}_{-0.09} M_\odot$ neutron star mass, assuming small surface velocities. The HS model mass is very similar. Coincidentally this is close to the value quoted by \citetalias{linares2018peering}. Because WH and HS produce large decreases in $\chi^2$ over DH models and explain the otherwise arbitrary phase of maximum, we believe that such models, with their more reasonable mass, are strongly preferred. 

As for J1959, if highly supersonic surface motions are allowed, we can accommodate a somewhat smaller mass, just consistent with $2M_\odot$ at the $1\sigma$ level (although in this case the motion would need to be, implausibly, opposite to the bulk heat flow). Improved spectroscopy to measure or bound such unexpectedly large surface motion is certainly desirable.

\section{Conclusion}\label{conclusion}

We have examined global wind circulation in  `spider'-type companion-evaporating pulsars, finding that the light curves and radial velocity curves of these objects can be significantly affected by wind heat advection. By applying a simple energy transport model, we see that the resulting synthetic light curves can explain otherwise puzzling light curve maximum phase shifts and asymmetries. These effects are sensitive to the ratio of the radiative and advective times in the sub-photospheric heating zone. These winds are analogous to those inferred for hot Jupiters, albeit at higher companion temperatures. However, in the MSP case, we have no optically bright primary star to contend with; the optical signal comes purely from the wind-affected object. This means that detailed photometry and spectroscopy allow much more sensitive probes of the wind dynamics than can be obtained for the hot Jupiters.

Global winds (and likely any model with large scale flows) move heat away from the sub-pulsar point. This moves the center of light closer to the center of mass, reducing the correction from the former (observed) quantity. In addition, the heat flow tends to broaden the light curve peak from direct heating models. This allows models with higher inclination $i$ to fit the photometric data. Together, these two effects tend to decrease the masses from those inferred from direct heating models. They also naturally lower the required $T_N$, which is surprisingly high for many BWs. Thus, these effects are quite important to consider when measuring spider pulsar masses. Importantly, global wind flow introduces characteristic asymmetries in the light curves, radial velocities, and line width variation, so that high-quality observations can provide excellent tests for these effects. 

We have applied our model to a BW  (PSR  J1959+2048) and a RB (PSR J2215+5135), both of which show strong pulsar irradiation, and compared our model with a heuristic model of a heated star spot. For the BW J1959, the WH model is significantly preferred, by both the photometric and spectroscopic fits; the fit gives us useful constraints on the wind parameters ($\epsilon$ and $\theta_c$). For the RB J2215, while the WH improves greatly over a direct heating fit, a simple star spot model is even better. Since RB companions are low mass fully-convective stars, we expect that the rapid rotation will drive powerful dynamos, giving rise to strong magnetic fields and, possibly, poles heated by particle precipitation. Indeed, light curve asymmetries attributable to star spot seem relatively common for RB. We can speculate that BWs have, in some cases, weaker convection and surface fields. With weaker fields, large-scale heat transport by winds may be a plausible means of introducing asymmetry. More examples should be studied, and more detailed light curves and spectra are needed to infer the physical origin of the asymmetric heat distribution (as described above) before one should make a generalization. 

Applying the revised heating pattern from the WH photometry models also allows improved fits to the radial velocity data. As expected, the WH model implies smaller neutron star masses: $2.18\pm0.09M_\odot$ for J1959 and $2.28^{+0.10}_{-0.09}M_\odot$ for J2215. For the latter, the HS fit is even better giving a very similar $2.24\pm 0.09\,M_\odot$; WH and HS fits reduce the mass from a very large, likely unphysical DH value. However, while decreased, these masses are still interestingly large. Both exceed $2M_\odot$ at the $2\sigma$ level ($2.7\sigma$ level for J2215), and exceed the present largest mass from Shapiro delay in pulsar timing ($2.14 ^{+0.10}_{-0.09}M_\odot$ for J0740+6620, \citealt{cromartie2019relativistic}). At face value, our measurements can thus be very important for constraining the equation of state at supernuclear densities.

Since details of the surface heating are important in computing spider masses from optical photometry and spectroscopy, such masses naturally engender less confidence than those directly from pulsar timing. But as the sophistication of our companion modeling improves, and physically motivated models provide increasingly accurate descriptions of the optical data, confidence in these results may increase. Global winds are an excellent example, with good physical motivation for heat transport from the sub-pulsar point and rather direct analogs observed in hot Jupiters. Our model, amenable to direct data fits, can produce a marked improvement in describing the photometric data. Since the principal effect is from the revised heating pattern, any model that similarly improves the match to the photometric data (e.g. an arbitrary collection of star spots, or heating by IBS flux) will result in similar values for the binary parameters.

Our study also points to the importance of epoch variability in these binaries. In particular, it seems that the veiling flux can change from year to year. There is also evidence in RB (e.g.  \citealt{2016ApJ...833L..12V}) that the heating asymmetries can vary; this is naturally expected if we attribute this to magnetic poles in a rapidly evolving dynamo. However, we might also expect global wind patterns to be highly turbulent and variable. This means that for neutron star mass measurements it is safest to extract light curve constraints on the detailed heating pattern at the same epoch as the critical radial velocity measurements. Carefully flux calibrated spectra are, at minimum, needed to check that the light curve is in a similar state.

However, at this point the largest model mass uncertainty in our study is the absence or presence of supersonic photospheric flows. At the sub-pulsar point sound speed (8\,km/s for J1959 and 10\,km/s for J2215) the corresponding mass shifts are $\pm0.07M_\odot$ and $\pm 0.05M_\odot$, respectively. These are important for precision values but do not dramatically change our conclusions. These kinematic effects can be probed by higher resolution optical spectroscopy. Already, our improved companion modeling, including the effect of global winds, bolsters confidence that these spider binaries are among the most massive neutron stars known. Additional spectroscopy can put the physical assumptions of these models to the test, checking the high masses and better probing the nature of the wind- or spot-induced light curve asymmetries.
\bigskip

We thank Alex Filippenko and colleagues for continued collaboration in obtaining optical data on spider pulsars which motivated much of the modeling work above. Questions raised by the anonymous referee also spurred important clarifications in the discussion. This work was supported in part by grant 80NSSC17K0024.

\appendix

We have noted above differences with the model fits of LSC18 for PSR J2215+5135. Since we start with very similar data sets, we have attempted to trace the origin of these discrepant fit values. The first is the modest difference in the fit inclination, our $i\sim 58.1^\circ$ versus their $i\sim 63.9^\circ$. To understand this discrepancy, we attempted to duplicate \citetalias{linares2018peering}'s analysis by combining the IAC-80, N1 and N2 magnitudes (with the arbitrary amplitude shifts they allow between all data sets and filters) into a single light curve for each color. Fitting these combined light curves, we obtained a best-fit inclination of $i\sim 62.0\pm 2.5^\circ$, consistent with the value in \citetalias{linares2018peering}. The remaining difference may be due to the fact that they binned the photometry into 37 ($g^\prime,i^\prime$) and 28 ($r^\prime$) phase bins before fitting. We note that our phase-shifted unbinned DH fit has fewer parameters and is statistically preferred (but is in any case superseded by our WH/HS fits).

For the velocities our corrections from $K_o$ to $K_c$ are quite different, especially for the MgI case. \citetalias{linares2018peering} in fact find $K_c < K_{\rm MgI}$ for this case. We do not find this to be consistent with the heating pattern, which would require the MgI line flux to be very strongly concentrated to the back (night) side. At any given phase the absorption line radial velocity is computed by a weighted sum over the $s$ tiles of the companion surface
$$
v=\sum_{s}(f_s v_s\times EW_s)/\sum_{s}(f_s\times EW_s),
$$
where $f_s$ is the surface thermal flux directed at the observer and, as emphasized by \citetalias{linares2018peering}, the absorption line equivalent width $EW_s$ are a function of the local surface temperature. For consistency, we use the particular $EW(T)$ relationships plotted by \citetalias{linares2018peering}. While the MgI $EW$ is indeed largest at the low night side temperatures, the larger day side fluxes $f_s$ keep the absorption line centroids from night side domination. To reproduce the \citetalias{linares2018peering} $K_c$ values, we had to modify the weighting. The simplest possibility is to drop the $f_s$ terms (weight the lines {\it only} by equivalent width). Alternatively, one might erroneously include the veiling flux in $f_s$ so that the effective night side flux was a larger fraction of the day flux, while erroneously having a large absorption line equivalent width. However, to induce backside dominance one requires the veiling flux to be $>300\mu$Jy, substantially larger even than \citetalias{linares2018peering}'s suggested value. Finally, if temperatures on the front side are large enough one can indeed reduce $EW_s$ even at the terminator so that the backside dominates. In practice, we found that this required pulsar heating fluxes $L_P>5\times 10^{35}{\rm erg\,s^{-1}}$, again even larger than found by \citetalias{linares2018peering} (and much larger than the total spindown luminosity); it is in any case difficult to see how such high $T$ could be consistent with the observed light curve colors. 

We speculate that they may have made one of the errors noted above -- for example ignoring $f_s$ in the weights then gives $K_c{\rm (MgI)}=395\pm 8\,{\rm km\,s^{-1}}$ for their observed MgI velocities. In this case, \citetalias{linares2018peering}'s Balmer $K_c{\rm (B)}=429\pm 9\,{\rm km\,s^{-1}}$ would be $4\sigma$ away from their MgI value; with combined fitting giving an intermediate value of $K_c=411.2\pm 6.0\,{\rm km\,s^{-1}}$, used in their mass estimate. Similarly the large $L_p$ error gives $K_c{\rm (MgI)} = 400\pm 8\,{\rm km\,s^{-1}}$, still $4\sigma$ from the Balmer value of $432\pm 9\,{\rm km\,s^{-1}}$ (combining to $K_c = 414\pm 6\,{\rm km\,s^{-1}}$). In contrast, with the correct weighting we get $K_c {\rm (MgI)} = 430\pm 9\,{\rm km\,s^{-1}}$ and $K_c {\rm (B)} = 448\pm 10\,{\rm km\,s^{-1}}$ in reasonable agreement, combining to $K_c=436\pm 7\, {\rm km\, s^{-1}}$; results with our own spectral measurements are quite similar and consistent. However, with this large $K_c$ and the best-fit $i$ the correctly treated DH model predicts an improbable $M_{\rm NS} = 3.05\pm 0.18\,M_\odot$. Since \citetalias{linares2018peering}'s observed radial velocity amplitudes and best fit $i$ are similar, we believe that \citetalias{linares2018peering} should have also found a large mass. Thus, the improved $\chi^2$ for light curve fitting, the explanation for the very significant phase shift and the more modest mass lead us to prefer the HS (or WH) model. In light of the apparent inconsistencies in their application of a DH model, the similarity of our HS/WH mass to that of \citetalias{linares2018peering} is a coincidence.

\bibliographystyle{aasjournal}
\bibliography{mainpaper}
\end{document}